# Precise Layer-Dependent Electronic Structure of MBE-Grown PtSe$_2$

Lei Zhang,[†,‡] Tong Yang,[‡,*] Muhammad Fauzi Sahdan,[‡,#] Arramel,[‡] Wenshuo Xu,[†,‡] Kaijian Xing,[%] Yuan Ping Feng,[‡,#] Wenjing Zhang, Zhuo Wang,[†,*] and Andrew T. S. Wee[‡,#,*]

[†]SZU-NUS Collaborative Innovation Center for Optoelectronic Science & Technology, International Collaborative Laboratory of 2D Materials for Optoelectronics Science and Technology of Ministry of Education, Institute of Microscale Optoelectronics, Shenzhen University, Shenzhen 518060, China.

[‡]Department of Physics, National University of Singapore, 2 Science Drive 3, Singapore 117542, Singapore

[#]Centre for Advanced 2D Materials and Graphene Research Centre, National University of Singapore, 6 Science Drive 2, Singapore 117546, Singapore

[%]ARC Centre for Future Low Energy Electronics Technologies, School of Physics and Astronomy, Monash University, Clayton, VIC, Australia.

**ABSTRACT:** Two-dimensional (2D) platinum diselenide (PtSe$_2$) has received significant attention for 2D transistor applications due to its high mobility. Here, using molecular beam epitaxy, we investigate the growth of 2D PtSe$_2$ on highly oriented pyrolytic graphite (HOPG) and unveil their electronic properties via X-ray photoelectron spectroscopy, Raman spectra, and scanning tunnelling microscopy/spectroscopy as well as density functional theory (DFT) calculations. PtSe$_2$ adopts a layer-by-layer growth mode on HOPG and shows a decreasing band gap with increasing layer number. For the layer numbers from one to four, PtSe$_2$ has band gaps of 2.0 ± 0.1, 1.1 ± 0.1, 0.6 ± 0.1 and 0.20 ± 0.1 eV, respectively, and becomes semimetal from the fifth layer. DFT calculations reproduce the layer-dependent evolution of both the band gap and band edges, suggest an indirect band-gap structure, and elucidate the underlying physics at the atomic level.

As an emerging two-dimensional (2D) material, platinum diselenide (PtSe$_2$) has attracted enormous attention in recent years in both fundamental science and practical applications due to its novel physical and chemical properties, such as facile synthesis, tunable band gap, high mobility and small electron effective mass, spin-layer locking, strong interlayer coupling, defect-induced layer-modulated magnetism and robust stability in air. These properties make it a promising candidate in various fields like electronics,[1–5] optoelectronics,[6–13] spintronics,[14] catalysis,[15] micro-electromechanics,[16] and sensing.[6,17,18] Monolayer or few-layer PtSe$_2$ can be synthesized by different methods, such as direct selenization of Pt films at a low temperature (≤ 400 °C),[3,4,6,8,19] which makes it scalable and compatible with current silicon chip fabrication technology, molecular beam epitaxy (MBE),[16,20,21] chemical vapor deposition (CVD),[5,22] and chemical vapor transport (CVT).[1,23] While PtSe$_2$ is a semimetal in bulk,[23] it becomes a semiconductor when thinned down to a few layers, due to the quantum confinement effect. Its electronic structure has been studied by angle-resolved photoemission spectroscopy (ARPES), which reveals the semiconducting property of monolayer PtSe$_2$ with the top of its valence band located at 1.2 eV below the Fermi level.[17,20] Complemented by density functional theory (DFT) calculations under the localized-density approximation (DFT-LDA), monolayer PtSe$_2$ is determined to be an indirect-gap semiconductor (~1.2 eV).[17] Besides the first layer, PtSe$_2$ remains semiconducting at the thickness of bilayer with a significantly reduced gap of 0.21 eV and becomes semimetallic from the third layer, as predicted by DFT-LDA calculations.[17,24] However, the ARPES measurement only gives the area-average information of the band structure below the Fermi level and its results are subject to the crystal size and quality. For example, top layers always appear amid the growth of the first layer, making it challenging to measure the precise band structure of monolayer. On the other hand, different calculation methods have yielded different values of the band gap. For instance, the band gap of the monolayer PtSe$_2$ has been predicted in a wide range of 1.05 – 2.48 eV,[2,17,18,24–29] while that of the bilayer in the range of 0.21 – 0.80 eV.[2,17,24,25] The band gap is one of the most important electronic parameters, and is crucial for electronic applications. Yet so far, the band gaps of monolayer and few-layer PtSe$_2$ have been determined only by ARPES measurements complemented by DFT calculations. The precise evolution of its local, and not average band gap as a function of layer number, as well as a deep understanding of such evolution are still absent, addressing which is important for the applications of PtSe$_2$.

In this letter, we report the study of high-quality MBE-grown monolayer and few-layer PtSe$_2$ on highly oriented pyrolytic graphite (HOPG) by X-ray photoelectron spectroscopy (XPS), Raman spectra, and scanning tunnelling microscopy/spectroscopy (STM/STS) as well as DFT calculations. PtSe$_2$ adopts a layer-by-layer growth mode on HOPG and shows a decreasing band gap with increasing layer numbers. The local band gap values obtained in our STS experiment are 2.0 ± 0.1, 1.1 ± 0.1, 0.6 ± 0.1 and 0.20 ± 0.1 eV for 1 – 4 layers, respectively, and 0 eV for layer numbers ≥ 5. DFT calculations well reproduce such semiconductor-to-semimetal transition and further suggest an indirect band-gap structure. This work provides a systematic study of the local layer-dependent electronic structure of PtSe$_2$, and gains a deep insight into such evolution at the atomic level.

**RESULTS AND DISCUSSION**

PtSe$_2$ crystallizes in a trigonal (1T) crystal structure with $p\bar{3}m1$ space group (**Figure 1a**). The crystal structure consists of a sublayer of hexagonally arranged Pt atoms sandwiched between two Se sublayers with in-plane and out-of-plane lattice constants of ~3.75 Å and 5.6 Å, respectively. These parameters have been confirmed by our STM measurements (see discussion below). Thin PtSe$_2$ films were grown by co-deposition of Pt and Se atoms onto HOPG at 280 °C (see schematic in **Figure 1b**). The growth of PtSe$_2$ films was monitored by XPS and Raman spectra. **Figure 1c** shows the XPS of Pt 4*f* and Se 3*d* core levels. All spectra are normalized to the same height to highlight the line shape evolution. The Pt 4*f* peaks at binding energies of 73.0 and 76.2 eV belong to the Pt$^{4+}$ in PtSe$_2$, and the Se 3*d* peak at a binding energy of 54.5 eV corresponds to Se$^{2-}$ in PtSe$_2$, consistent with previous reports.[17,20] As a comparison, the XPS of clean Pt (Pt$^0$) and Se (Se$^0$) films are also shown. In contrast to the blue shift of both Pt 4*f* and Se 3*d* core levels of MBE-grown PtSe$_2$ on bilayer graphene/6H-SiC(0001),[20] no obvious change of the line shape and peak positions with increasing thickness was observed in our XPS results, suggesting no charge transfer between the HOPG substrate and PtSe$_2$ films. **Figure 1d** shows the Raman spectra with varying thickness. Three peaks are clearly identified at ~ 175.0, 202.5 and 229.5 cm$^{-1}$, which correspond to the E$_g$, A$_{1g}$ and longitudinal optical (LO) Raman active modes, respectively.[20,25] The E$_g$ and A$_{1g}$ modes are in-plane and out-of-plane vibrational modes of Se atoms moving in opposite directions, respectively. The E$_g$ mode shows a minor red shift with increasing film thickness, whereas the A$_{1g}$ mode remain unshifted, which may be attributed to the stacking-induced structural changes and long-range Coulombic interactions, in agreement with previous work and theoretical calculations.[25,30] The peak intensity of the A$_{1g}$ mode relative to that of the E$_g$ mode shows a significant increase in intensity with increasing film thickness, due to enhanced out-of-plane interactions from increased layer numbers.[30] The XPS and Raman characterization demonstrates the good crystalline quality of our MBE-grown PtSe$_2$.

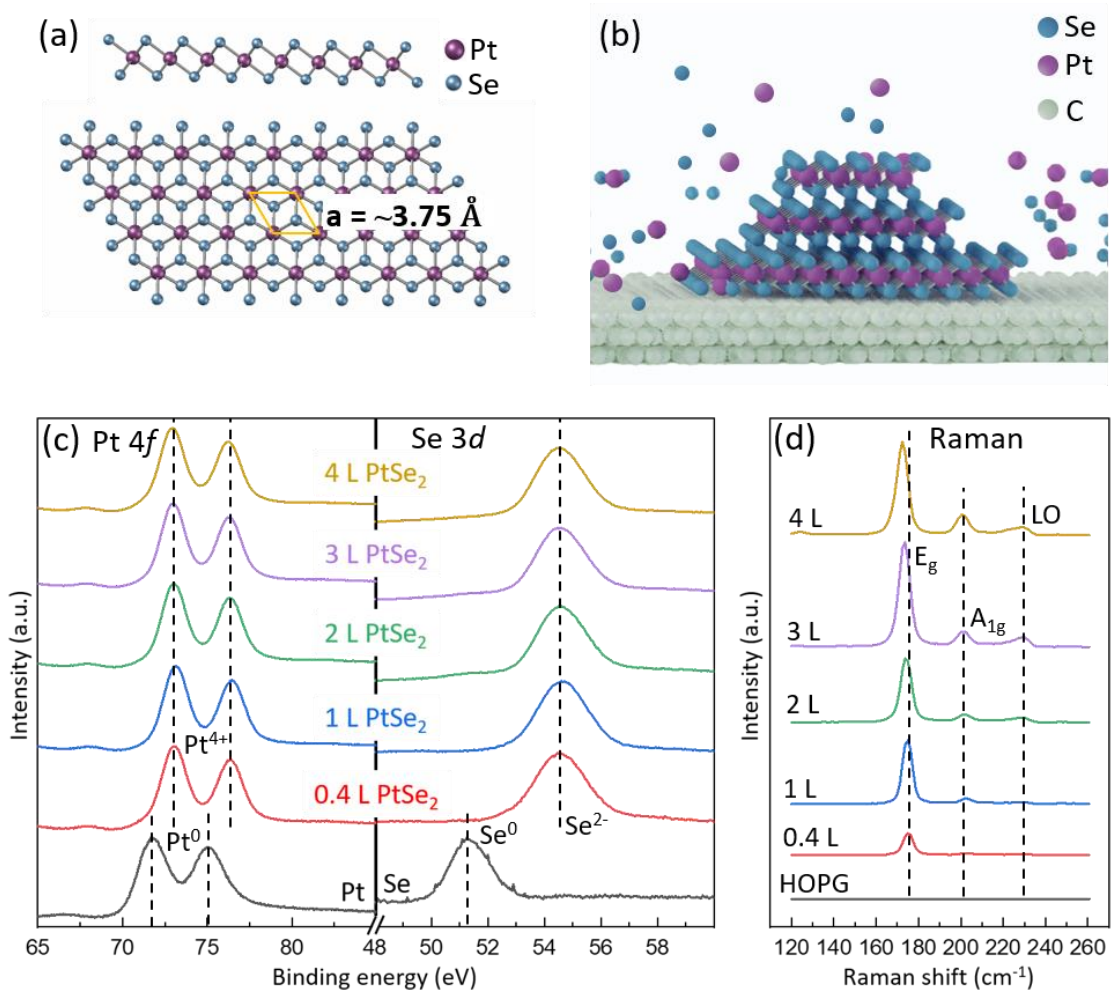

**Figure 1**. **Growth process, XPS and Raman spectra of PtSe$_2$ films.** (a) Side-view and top-view lattice structure of the 1T-PtSe$_2$. (b) Schematic of the fabrication of PtSe$_2$ thin films by co-deposition of Pt and Se atoms onto HOPG substrate. (c) XPS measurements of Pt 4$f$ and Se 3$d$ core levels demonstrating the formation of PtSe$_2$ at 280 °C. (d) Raman spectra showing the E$_g$ mode, A$_{1g}$ mode and LO mode.

**Figure 2a – d** present the STM topographic evolution of PtSe$_2$ on HOPG with increasing thickness. The second and third layers appear even at an extremely low coverage (**Figure 2a**), indicating a stronger interlayer interaction in PtSe$_2$ than that between the substrate and the first-layer PtSe$_2$. With increasing thickness, more and more top layers can be seen. The nearly hexagonal and triangular shapes of PtSe$_2$ imply the single crystalline nature of the samples, as shown in **Figure 1b – d**. The height profile along the white dashed line in Figure 1b indicates an interlayer distance of ~0.56 nm, as shown in **Figure 2e**. **Figure 2f** shows the atomic resolution

STM image of the first-layer PtSe$_2$, showing the hexagonal lattice of Se atoms in the topmost sublayer of the PtSe$_2$ sandwich-type structure and a lattice constant of ~3.75 Å. The atomic resolution images of the first six layers of PtSe$_2$ are shown in **Figure S1**. All these layers have the same lattice structure and similar lattice constants. The growth behavior revealed by our STM characterization implies that the influence of top layers in averaged ARPES data ought to be considered for measurements on MBE-grown monolayer PtSe$_2$.

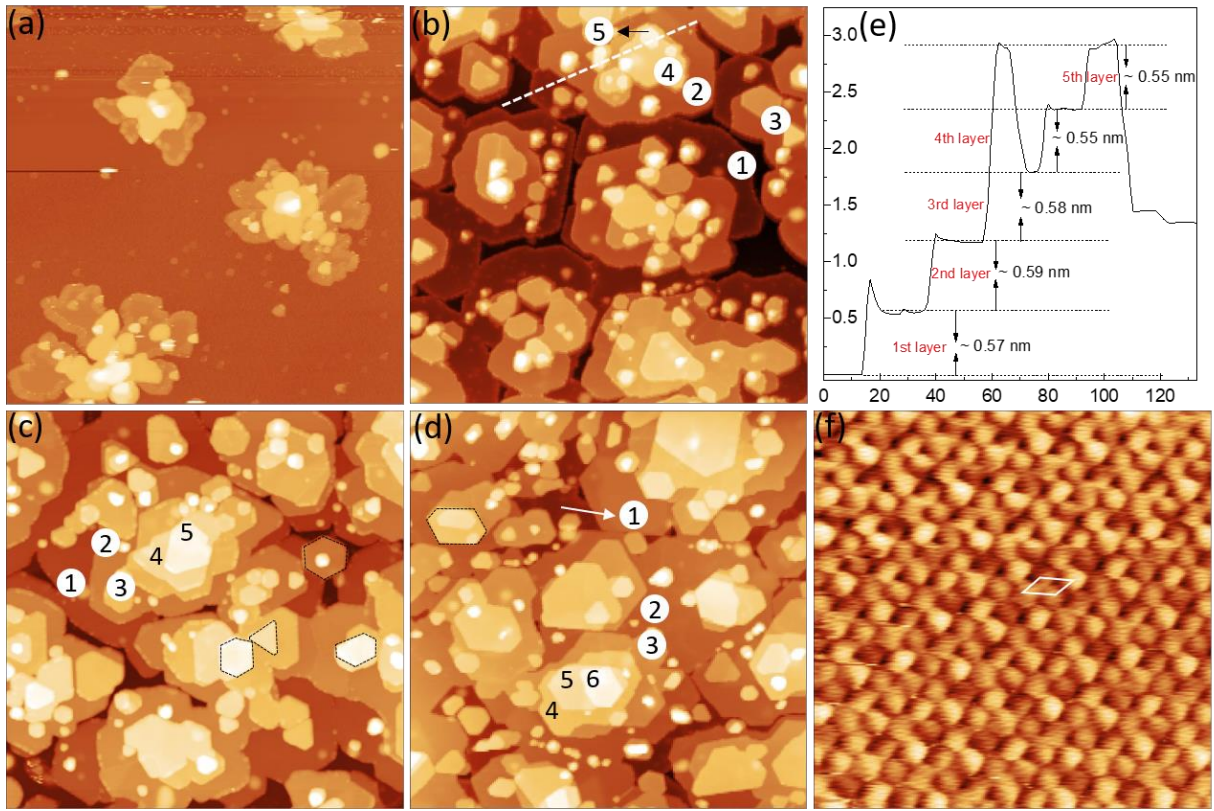

**Figure 2**. **Topographic evolution of PtSe$_2$ on HOPG.** (a) – (d): STM images of PtSe$_2$ on HOPG with increasing thickness. Layer numbers are indicated by numbers. (Size: 250 × 250 nm$^2$. Setpoints: a, 1.0 V, 10 pA; b, 1.7 V, 10 pA; c, –1.8 V, 14 pA; d, 1.5 V, 12 pA) (e): Height profile corresponding to the white dashed line in (b). (f): Atomic resolution STM image of the first-layer PtSe$_2$. (6 × 6 nm$^2$; Setpoint: 80 mV, 415 pA).

In view of the complex topology of few-layer PtSe$_2$ on HOPG, it is important to determine the local layer-dependent atomic-scale electronic structure. **Figure 3a** shows the layer-dependent STS. The first layer has a band gap of 2.00 ± 0.1 eV, and its valence band maximum (VBM) is located at 1.47 eV below the Fermi level. In references [17] and [20], the VBM was observed to be at 1.2 eV below the Fermi level by ARPES. Such difference from our observation is probably caused by the charge transfer between the substrate (for example, bilayer graphene/6H-SiC(0001) in reference [20]) and the first-layer PtSe$_2$, as suggested by the XPS measurement in reference [20], where they found a 0.24 eV blue shift in the Se 3$d$ and Pt 4$f$ core levels in the first layer compared to other layers. However, such shift is not observed in our XPS, suggesting no charge transfer between the first-layer PtSe$_2$ and the HOPG substrate. Taking into account this shift, the VBM position obtained in our STS (–1.47 eV) is consistent with these ARPES results (–1.2–0.24 = –1.44 eV). In contrast to previous DFT calculations on the second layer predicting a quite small band gap of 0.21 – 0.35 eV,[2,17,24] our STS shows that the second layer still has a sizable band gap of 1.10 ± 0.1 eV. The VBM of the second layer is found to be at 0.80 eV below the Fermi level, which is deeper than previous ARPES result.[20] This might be due to the imperfect layer-by-layer growth behavior as revealed by our STM, which influence the precise determination of its VBM by ARPES. With increasing layer numbers, the band gap decreases remarkably to 0.60 ± 0.1 eV for the third layer and 0.20 ± 0.1 eV for the fourth layer, and vanishes from the fifth layer. Compared to the shift of the conduction band minimum (CBM) with increasing layer number, the shift of the VBM is larger, which is ascribed to the higher contribution of the Se $p_z$ orbital to top valence bands, as discussed below.

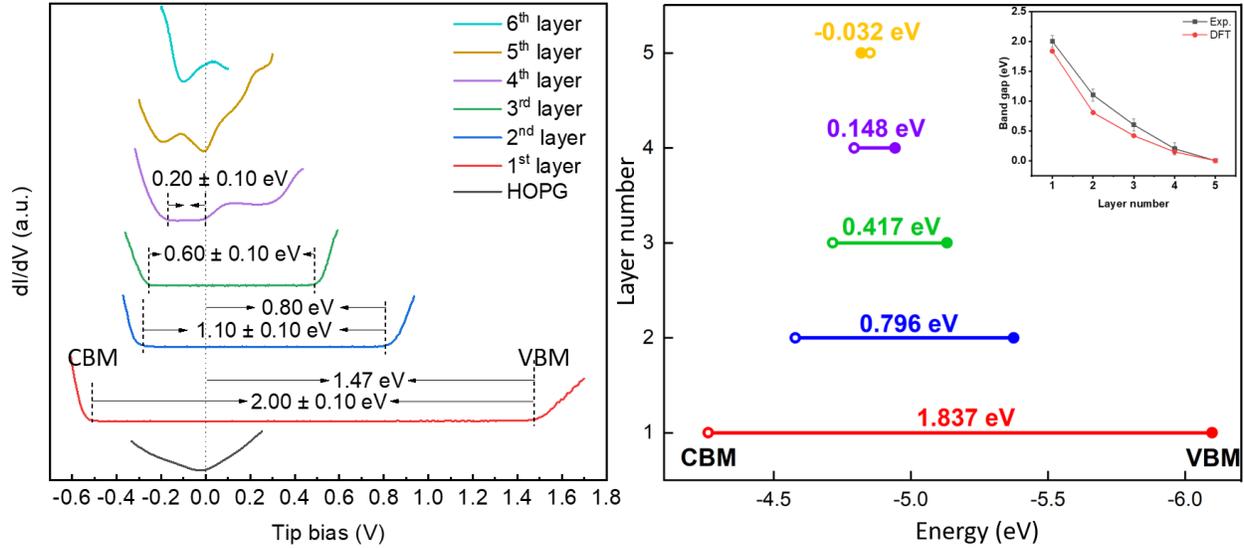

**Figure 3**. **STS and band edges of PtSe$_2$ on HOPG.** (a): Layer-dependent STS showing the evolution from semiconductor to semimetal with layer numbers from one to six. (b): The calculated layer-dependent band gaps and band edges (VBM and CBM) of PtSe$_2$ at the HSE06 level. The solid and empty circles denote the VBM and the CBM, respectively. Inset: Comparison of thickness-dependent band gaps obtained by STS (black squares) and DFT calculations (red dots).

The semiconductor-to-semimetal transition is further supported by our DFT calculations. To accurately describe the interlayer interaction, optB88-vdW functional has been adopted to optimize the PtSe$_2$ structure. **Table S1** shows the optimized lattice constants of bulk PtSe$_2$, which are in good agreement with previous experimental reports.[23] As discussed above, however, previous DFT calculations based on either the optB88-vdW functional or LDA predicted a much smaller band gap for PtSe$_2$.[2,17,24] Here, the Heyd-Scuseria-Ernzerhof hybrid functional (HSE06) was utilized to obtain a more accurate electronic band structure. The calculated band structures of monolayer to five-layer PtSe$_2$ as well as the bulk PtSe$_2$ are visualized in **Figure S2** with the evolution of the band gap size and edges (VBM and CBM) with the layer number summarized in **Figure 3b**. DFT calculations reveal the indirect band-gap nature of monolayer and few-layer PtSe$_2$ (**Figure S2**). The calculated layer-dependent band gaps are very similar to those derived from our STS results (Inset of **Figure 3b**). In particular, the semiconductor-to-semimetal transition observed from the fourth layer to fifth layer is well captured by our calculations. With regard to the band edges, **Figure 3b** shows that the VBM (solid circles) shifts more significantly than the CBM (empty circles) upon stacking PtSe$_2$, which is also consistent with our STS results (**Figure 3a**).

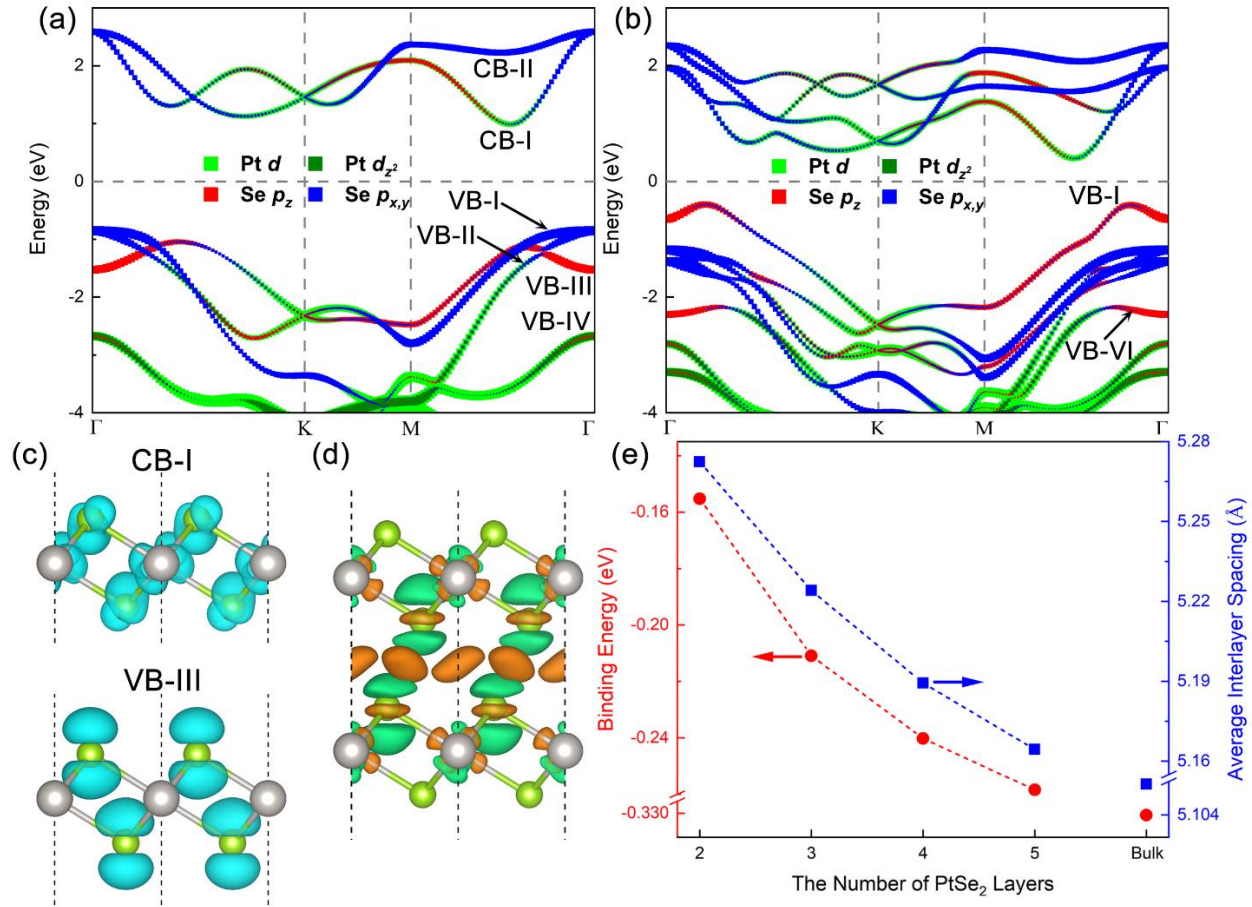

**Figure 4. DFT calculations.** (a) and (b): The projected band structures of monolayer and bilayer PtSe$_2$, respectively. VB-n (CB-n) denotes the *n*-th highest occupied (lowest unoccupied) band around the Γ point. (c): The visualized wavefunction of (upper panel) CB-I at CBM and (lower panel) VB-III at the Γ point in monolayer PtSe$_2$. (d): The charge redistribution upon stacking two PtSe$_2$ layers. Orange (green) represents charge accumulation (depletion). (e): The interlayer binding energy and average interlayer spacing as functions of the layer number.

To shed light on the dependence of the band gap size and edges on the layer number, we first concentrate on the electronic structures of monolayer and bilayer PtSe$_2$, as shown in **Figure 4a** and **b**. Wherein, we label the *n*-th highest occupied (lowest unoccupied) band around the Γ point as VB-*n* (CB-*n*). For the monolayer PtSe$_2$, the VB-I is double degenerate at the Γ point, which is dominated by the in-plane Se $p_x$ and $p_y$ orbitals. On the other hand, the out-of-plane Se $p_z$ orbital forms VB-III, whereas the Pt $d_{z^2}$ orbital slightly hybridizes with the Se $p_z$ orbital and leads to a deep VB-IV. As for the conduction band, the CB-I is nearly halfway between the M and Γ points and is from the hybridization between the Pt $d$ and Se $p$ orbitals. When two layers of PtSe$_2$ are stacked together, the small interlayer wavefunction overlap of the original Se $p_{x,y}$-dominated VB-I and VB-II leads to a small energy splitting, as manifested in **Figure 4b**. On the contrary, the considerable out-of-plane extension of the Se $p_z$ orbital induces a strong interlayer hybridization between them, thereby leading to a large energy splitting of the original VB-III (1.653 eV). As shown in **Figure 4b**, such splitting is so large that the anti-bonding Se $p_z$ state becomes the VB-I of the bilayer PtSe$_2$. For the conduction band edge, **Figure 4c** shows that the wavefunction of CB-I at CBM is more localized within the triatomic layer of PtSe$_2$, compared to that of VB-III. Therefore, a smaller energy splitting of CB-I is observed in **Figure 4b**. We also calculated the charge redistribution upon the formation of bilayer PtSe$_2$, as displayed in **Figure 4d**. The $p_z$-like charge depletion (green) around the inner Se atoms further implies that the hybridization between interlayer Se $p_z$ orbitals makes a major contribution to the interlayer interaction. Based on the above analysis, the reduced band gap can be mainly attributed to the strong hybridization of interlayer Se $p_z$ orbitals upon the formation of bilayer PtSe$_2$. It is worth noting that S $p_z$ orbital has been reported to play the similar role in reducing the band gap of bilayer PtS$_2$.[31] Meanwhile, the larger shift of VBM is ascribed to the higher contribution of Se $p_z$ orbital to the top valence bands. As more PtSe$_2$ layers are stacked together, **Figure 4e** shows that both the interlayer binding energy (see METHODS) and the average interlayer spacing continue decreasing, indicating a stronger interlayer interaction. As a result, the band gap of PtSe$_2$ continues decreasing with increasing layer number until the semiconductor-to-semimetal transition is reached at the layer number of five. Our DFT calculations well reproduce the experimental observations and give a deep understanding to the physics of the layer-dependent evolution of the electronic structure of PtSe$_2$ at the atomic level. The stacked structure and layer-dependent band gap of PtSe$_2$ may enable broad-band optical absorption and photoresponse in applications such as field effect transistor.

**CONCLUSIONS**

We have systematically studied the local layer-dependent electronic structures of PtSe$_2$ by XPS, Raman spectra, STM/STS and DFT calculations. It is revealed that monolayer and few-layer PtSe$_2$ is an indirect semiconductor with its band gap decreasing with increasing layer numbers. The precise band gaps obtained by STS are 2.00 ± 0.10, 1.10 ± 0.10, 0.60 ± 0.10 and 0.20 ± 0.10 eV for the first to the fourth layers, respectively, and 0 eV from the fifth layer. The reducing band gap size and large energy shift of the valence band edge with increasing layer number are mainly attributed to the strong hybridization of interlayer Se $p_z$ orbitals and the high Se-$p_z$ composition in top valence bands. This work not only determines the important electronic parameter, *i.e.* layer-dependent band gaps, but also gives a deep understanding to the physics of the evolution of the electronic structure of PtSe$_2$. The widely tunable layer-dependent bandgap may enable PtSe$_2$ to expand its photoresponsivity spectrum into the infrared range, as well as facilitate electronic device applications.

**METHODS**

**Sample Preparation**. PtSe$_2$ was grown on HOPG substrates by MBE in an ultrahigh vacuum (UHV) chamber with a base pressure of ~2 × 10$^{-9}$ mbar. Before growth, the freshly cleaved substrate was annealed at 720 K for 1 h. Se (Sigma, 99.99%) was evaporated from a Knudsen cell at 470 K, while Pt (ESPI Metals, 99.995%) was evaporated from an electron-beam evaporator with its evaporation rate controlled by the flux current. During growth, the substrate was kept at 555 K, a temperature higher than the sublimation temperature of Se, and the whole growth process was maintained under Se-rich conditions to ensure sufficient Se reacting with Pt. After growth, the sample was cooled to 320 K and capped with Se for ex-situ transfer to other UHV systems for subsequent STM/STS, XPS and Raman characterizations. Prior to these measurements, the Se-capping layer was desorbed at 520 K for 40 min in the preparation chamber of the respective systems.

**STM/STS Characterizations**. STM/STS measurements were carried out in a multi-chamber UHV system housing an Omicron LT-STM system interfaced to a Nanonis controller at 77 K. The base pressure was better than $10^{-10}$ mbar. A chemically etched tungsten tip was used, and the bias was applied to the STM tip. STM images were recorded in constant-current mode.

**XPS Characterization**. XPS experiments were performed in a VG ESCA Lab system with a typical base pressure in the range of $\sim 10^{-10}$ mbar and an X-ray source of magnesium Kα with a typical excitation energy output of 1254 eV. All the reported photoemission spectra here were calibrated by reference to the Ag $3d_{5/2}$ peak position (centered at 368.10 eV) of the silver polycrystalline foil cleaned via standard argon ion sputtering treatment. The errors of all the given values of the binding energies are estimated to be ±0.05 eV.

**Raman Characterization**. Raman spectra were captured by a commercial WITec Alpha 300 R Raman system. The continuous wave (CW) laser wavelength is 532 nm and the spot diameter is 0.5 μm using a 100× objective lens. A laser power of 0.65 mW, integration time of 1 s, and accumulation times of 2 were adopted for Raman spectrum measurement.

**DFT Calculations**. All density functional theory (DFT) calculations were carried out using the Vienna Ab initio Simulation Package (VASP).[32,33] The projector augmented-wave (PAW) method and the optB88-vdW functional were adopted to describe the core-electron interaction and the exchange-correlation interaction, respectively.[34–39] Compared with the Perdew-Burke-Ernzerhof parametrized generalized gradient approximation (PBE-GGA)[40] and the DFT-D3 method,[41] the optB88-vdW functional turns out to more accurately account for the interlayer interaction of $PtSe_2$, as shown in **Table S1**. It is noted that we switched to the Heyd-Scuseria-Ernzerhof hybrid functional (HSE06) to obtain a more accurate electronic band structure of $PtSe_2$.[42] The electronic wavefunction was expanded using a plane-wave basis with the cut-off energy of 450 eV. A Γ-centered $k$-mesh of $16 \times 16 \times 12$ and $16 \times 16 \times 1$ was adopted to sample the first Brillouin zone of multilayer and bulk $PtSe_2$, respectively. For multilayer $PtSe_2$, a vacuum layer of around 20 Å was inserted in the direction perpendicular to the basal plane in order to alleviate the spurious interlayer interaction. Both the lattice constants and the atomic positions were fully optimized until the force acting on each atom is less than 0.01 eV/Å. And the convergence criterion of the total energy calculation was set to $1\times10^{-5}$ eV. The interlayer binding energy of $n$-layer $PtSe_2$ is defined as $E_{\text{bind}}(n - \text{PtSe}_2) = \frac{E(n-\text{PtSe}_2)}{n} - E(\text{PtSe}_2)$, where $E(n - \text{PtSe}_2)$ and $E(\text{PtSe}_2)$ are respectively the total energies of $n$-layer and monolayer $PtSe_2$.

## ASSOCIATED CONTENT

**Supporting Information**.

Layer-dependent atomic resolution STM images; Optimized lattice constants of the bulk $PtSe_2$; Calculated layer-dependent band structures of $PtSe_2$. (PDF)

## AUTHOR INFORMATION

**Corresponding Authors**

*Andrew T. S. Wee: phyweets@nus.edu.sg

*Zhuo Wang: wzhuo@szu.edu.cn

*Tong Yang: yangtong@u.nus.edu

**Author Contributions**

The manuscript was written through contributions of all authors. All authors have given approval to the final version of the manuscript.

## ACKNOWLEDGMENT

Z.W. acknowledges financial support from National Natural Science Foundation of China (61805159), the Guangdong Natural Science Funds (2019A1515011007), Shenzhen Peacock Plan (827-000473) and Natural Science Foundation of SZU (2019015).

# Supporting Information

# Precise Layer-Dependent Electronic Structure of MBE-Grown PtSe$_2$

*Lei Zhang,* [†,‡] *Tong Yang,* [‡,*] *Muhammad Fauzi Sahdan,* [‡,#] *Arramel,* [‡] *Wenshuo Xu,* [†,‡] *Kaijian Xing,* [%] *Yuan Ping Feng,* [‡,#] *Wenjing Zhang, Zhuo Wang,* [†,*] *and Andrew T. S. Wee* [‡,#,*]

[†]SZU-NUS Collaborative Innovation Center for Optoelectronic Science & Technology, International Collaborative Laboratory of 2D Materials for Optoelectronics Science and Technology of Ministry of Education, Institute of Microscale Optoelectronics, Shenzhen University, Shenzhen 518060, China.

[‡]Department of Physics, National University of Singapore, 2 Science Drive 3, Singapore 117542, Singapore

[#]Centre for Advanced 2D Materials and Graphene Research Centre, National University of Singapore, 6 Science Drive 2, Singapore 117546, Singapore

[%]ARC Centre for Future Low Energy Electronics Technologies, School of Physics and Astronomy, Monash University, Clayton, VIC, Australia.

# 1. Layer-dependent atomic resolution STM images.

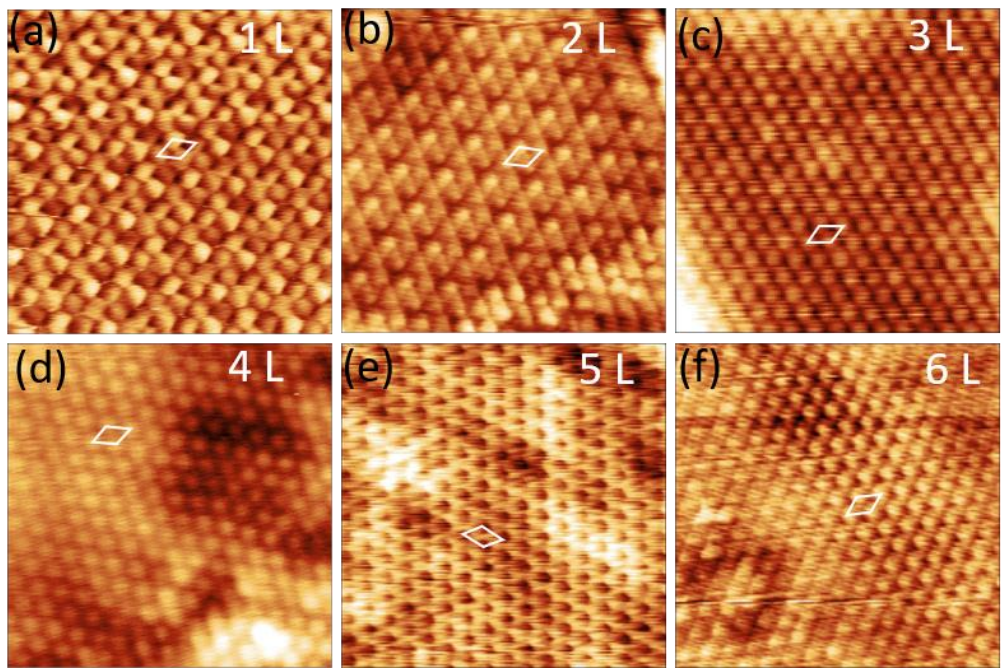

**Figure S1**. (a) – (f): Atomic resolution STM images of PtSe$_2$ on HOPG with layer numbers from one to six, showing the same lattice structure and similar lattice constants. (6 × 6 nm$^2$, setpoints: a, 80 mV, 440 pA; b, 200 mV, 206 pA; c, 300 mV, 329 pA; d, 80 mV, 415 pA; e, 50 mV, 496 pA; f, 150 mV, 261 pA).

## 2. Optimized lattice constants of the bulk PtSe$_2$.

**Table S1**: The optimized lattice constants of bulk PtSe$_2$ at the theory level of PBE-GGA, PBE-GGA with the DFT-D3 method and optB88-vdW, respectively. The experimental lattice constants are also tabulated.

|  | PBE-GGA | PBE-GGA w/ DFT-D3 | optB88-vdW | Experiments[1] |
|---|---|---|---|---|
| $a = b$ (Å) | 3.759 | 3.775 | 3.803 | 3.727 |
| $c$ (Å) | 6.121 | 4.802 | 5.110 | 5.07±0.01 |

## 3. Calculated layer-dependent band structures of PtSe₂.

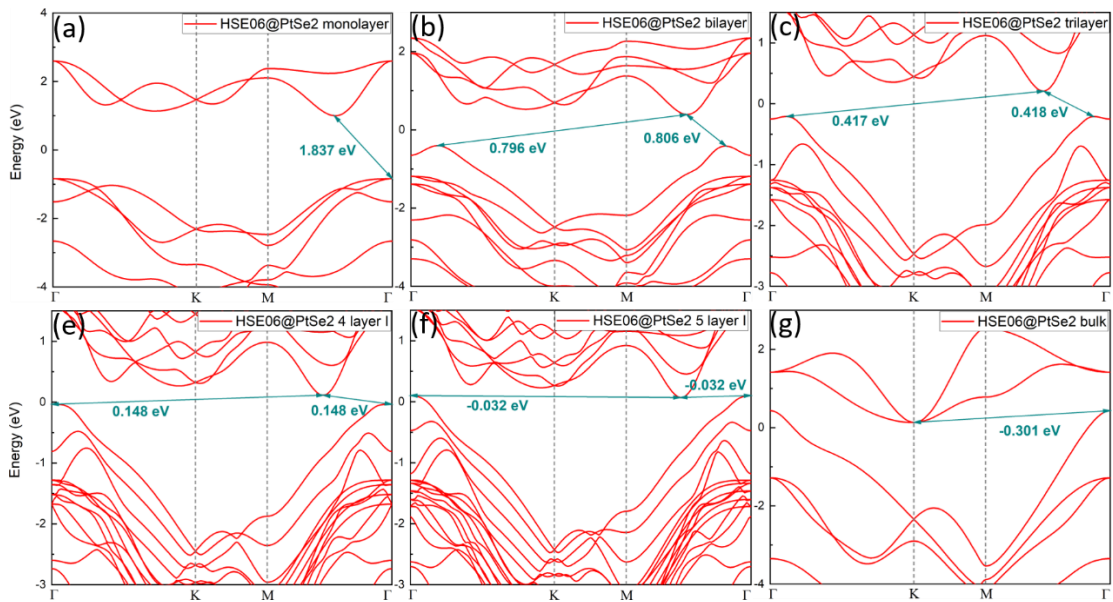

**Figure S2**. The calculated electronic band structures of PtSe$_2$ with layer number from one to five and bulk PtSe$_2$ at the HSE06 level.